\documentstyle[12pt,epsf]{article}

\title{%
\hfill
 \parbox{5cm}{%
 }\\
\vspace{5ex}
  CP and T violation in long baseline experiments with low energy
neutrino from muon storage ring
\vspace{5ex}
}

\author{
 Masafumi Koike%
\thanks{e-mail address: {\tt koike@icrr.u-tokyo.ac.jp}}\\%
{\footnotesize \it%
 Institute for Cosmic Ray Research, University of Tokyo,
 Midori-cho, Tanashi, Tokyo 188-8502, Japan%
}
\\
and
\\
Joe Sato%
\thanks{e-mail address: {\tt joe@hep-th.phys.u-tokyo.ac.jp}}\\%
 {\footnotesize \it%
   Department of Physics, University of Tokyo, Hongo, Bunkyo-ku, Tokyo
   113-0033, Japan%
 }%
}

\date{}

\begin{document}

\maketitle
\abstract{%
   Stimulated by the idea of PRISM, a very high intensity muon ring
   with rather low energy, we consider possibilities of observing
   CP-violation effects in neutrino oscillation experiments.  More
   than 10\% of CP-violation effect can be seen within the
   experimentally allowed region.  Destructive sum of matter effect and
   CP-violation effect can be avoided with use of initial $\nu_{\rm
   e}$ beam.  We finally show that the experiment with (a few)
   $\times$ 100 MeV of neutrino energy and (a few) $\times$ 100 km of
   baseline length, which is considered in this paper, is particularly
   suitable for a search of CP violation in view of statistical error.%
}

\section{Introduction}

Many experiments and observations have shown evidences for neutrino
oscillation one after another.  The solar neutrino deficit has long
been observed\cite{Ga1,Ga2,Kam,Cl,SolSK}.  The atmospheric neutrino
anomaly has been found\cite{AtmKam,IMB,SOUDAN2,MACRO} and recently
almost confirmed by SuperKamiokande\cite{AtmSK}.  There is also
another suggestion given by LSND\cite{LSND}.  All of them can be
understood by neutrino oscillation and hence indicates that neutrinos
are massive and there is a mixing in lepton
sector\cite{FukugitaYanagida}.

Since there is a mixing in lepton sector, it is quite natural to
imagine that there occurs CP violation in lepton sector.  Several
physicists have considered whether we may see CP-violation effect in
lepton sector through long baseline neutrino oscillation experiments. 
First it has been studied in the context of currently planed
experiment\cite{KP,Tanimoto,ArafuneJoe,AKS,MN,BGG} and recently in
the context of neutrino factory\cite{BGW,Tanimoto2,Romanino,GH}.

The use of neutrinos from muon beam has great advantages compared with
those from pion beam.  Neutrinos from $\mu^+$($\mu^-$) beam consist of
pure $\nu_{\rm e}$ and $\bar\nu_\mu$ ($\bar\nu_{\rm e}$ and $\nu_\mu$)
and will contain no contamination of other kinds of neutrinos.  Also
their energy distribution will be determined very well.  In addition
we can test T violation in long baseline experiments by using
(anti-)electron neutrino\cite{ArafuneJoe,AKS}.

Unfortunately those neutrinos have very high energy\cite{Geer}.  The
smaller mass scale of neutrino, determined by the solar neutrino
deficit, cannot be seen in most long baseline experiments since
CP-violation effect arise as three(or more)-generation
phenomena\cite{Cabibbo,BWP}, it is difficult to make CP violation
search using neutrinos from such muon beam.

We are, however, very lucky since we will have very intense muon
source with rather low energy, PRISM\cite{PRISM}.  It will located at
Tokai, Ibaraki Prefecture, about 50 km from KEK. Since the muons will
have energy less than 1 GeV, we can expect that we will have very
intense neutrino beam with energy less than 500 MeV. It will be very
suitable to explore CP violation in lepton sector with neutrino
oscillation experiments.  With such a low energy beam, we will be able
to detect neutrinos experimentally with good energy resolution. 
Stimulated by the possibility that we will have a low energy neutrino
source with very high intensity, we consider here how large
CP-violation effect we will see with such neutrino beam.

In this paper we will consider three active neutrinos without any
sterile one by attributing the solar neutrino deficit and atmospheric
neutrino anomaly to the neutrino oscillation.

\section{Oscillation probabilities and their approximated formulas}

First we derive approximated formulas\cite{AKS} of neutrino
oscillation \cite{FukugitaYanagida,BWP,BilenkyPetcov,Pakvasa} to
clarify our notation.

We assume three generations of neutrinos which have mass eigenvalues
$m_{i} (i=1, 2, 3)$ and MNS mixing matrix\cite{MNS} $U$ relating the
flavor eigenstates $\nu_{\alpha} (\alpha={\rm e}, \mu, \tau)$ and the
mass eigenstates in the vacuum $\nu\,'_{i} (i=1, 2, 3)$ as
\begin{equation}
  \nu_{\alpha} = U_{\alpha i} \nu\,'_{i}.
  \label{Udef}
\end{equation}
We parameterize $U$\cite{ChauKeung,KuoPantaleone,Toshev}
as

\begin{eqnarray}
& &
U
=
{\rm e}^{{\rm i} \psi \lambda_{7}} \Gamma {\rm e}^{{\rm i} 
\phi \lambda_{5}} {\rm e}^{{\rm i} \omega \lambda_{2}} \nonumber 
\\
&=&
\left(
\begin{array}{ccc}
  1 & 0 & 0  \\
  0 & c_{\psi} & s_{\psi} \\
  0 & -s_{\psi} & c_{\psi}
\end{array}
\right)
\left(
\begin{array}{ccc}
  1 & 0 & 0  \\
  0 & 1 & 0  \\
  0 & 0 & {\rm e}^{{\rm i} \delta}
\end{array}
\right)
\left(
\begin{array}{ccc}
  c_{\phi} & 0 &  s_{\phi} \\
  0 & 1 & 0  \\
  -s_{\phi} & 0 & c_{\phi}
\end{array}
\right)
\left(
\begin{array}{ccc}
  c_{\omega} & s_{\omega} & 0 \\
  -s_{\omega} & c_{\omega} & 0  \\
  0 & 0 & 1
\end{array}
\right)
\nonumber \\
&=&
\left(
\begin{array}{ccc}
   c_{\phi} c_{\omega} &
   c_{\phi} s_{\omega} &
   s_{\phi}
  \\
   -c_{\psi} s_{\omega}
   -s_{\psi} s_{\phi} c_{\omega} {\rm e}^{{\rm i} \delta} &
   c_{\psi} c_{\omega}
   -s_{\psi} s_{\phi} s_{\omega} {\rm e}^{{\rm i} \delta} &
   s_{\psi} c_{\phi} {\rm e}^{{\rm i} \delta}
  \\
   s_{\psi} s_{\omega}
   -c_{\psi} s_{\phi} c_{\omega} {\rm e}^{{\rm i} \delta} &
   -s_{\psi} c_{\omega}
   -c_{\psi} s_{\phi} s_{\omega} {\rm e}^{{\rm i} \delta} &
   c_{\psi} c_{\phi} {\rm e}^{{\rm i} \delta}
\end{array}
\right),
\label{UPar2}
\end{eqnarray}
where $c_{\psi} = \cos \psi, s_{\phi} = \sin \phi$, etc.

The evolution equation for the flavor eigenstate vector in the vacuum
is
\begin{equation}
 {\rm i} \frac{{\rm d} \nu}{{\rm d} x}
=
 \frac{1}{2 E}
 U {\rm diag} (0, \delta m^2_{21}, \delta m^2_{31})
 U^{\dagger}
 \nu.
 \label{VacEqnMotion}
\end{equation}
where $\delta m^2_{ij} = m^2_i - m^2_j$.

Similarly the evolution equation in matter is expressed as
\begin{equation}
 {\rm i} \frac{{\rm d} \nu}{{\rm d} x}
 = H \nu,
 \label{MatEqn}
\end{equation}
where
\begin{equation}
  H
  \equiv
  \frac{1}{2 E}
  \tilde U
  {\rm diag} (\tilde m^2_1, \tilde m^2_2, \tilde m^2_3)
  \tilde U^{\dagger},
 \label{Hdef}
\end{equation}
with a unitary mixing matrix $\tilde U$ and the effective mass squared
$\tilde m^{2}_{i}$'s $(i=1, 2, 3)$.  The matrix $\tilde U$ and the
masses $\tilde m_{i}$'s are determined by\cite{Wolf,MS,Earth}
\begin{equation}
\tilde U
\left(
\begin{array}{ccc}
  \tilde m^2_1 & & \\
  & \tilde m^2_2 & \\
  & & \tilde m^2_3
\end{array}
\right)
\tilde U^{\dagger}
=
U
\left(
\begin{array}{ccc}
  0 & & \\
  & \delta m^2_{21} & \\
  & & \delta m^2_{31}
\end{array}
\right)
U^{\dagger}
+
\left(
\begin{array}{ccc}
  a & & \\
  & 0 & \\
  & & 0
\end{array}
\right).
\label{MassMatrixInMatter}
\end{equation}
Here
\begin{equation}
 a \equiv 2 \sqrt{2} G_{\rm F} n_{\rm e} E \nonumber \\
   = 7.56 \times 10^{-5} {\rm eV^{2}} \cdot
       \left( \frac{\rho}{\rm g\,cm^{-3}} \right)
       \left( \frac{E}{\rm GeV} \right),
 \label{aDef}
\end{equation}
where $n_{\rm e}$ is the electron density and $\rho$ is the matter
density.  The solution of eq.(\ref{MatEqn}) is then
\begin{equation}
 \nu (x) = S(x) \nu(0)
 \label{nu(x)}
\end{equation}
with
\begin{equation}
 S \equiv {\rm T\, e}^{ -{\rm i} \int_0^x {\rm d} s H (s) }
 \label{Sdef}
\end{equation}
(T being the symbol for time ordering), giving the oscillation
probability for $\nu_{\alpha} \rightarrow \nu_{\beta} (\alpha, \beta =
{\rm e}, \mu, \tau)$ at distance $L$ as
\begin{eqnarray}
 P(\nu_{\alpha} \rightarrow \nu_{\beta}; E, L)
&=&
 \left| S_{\beta \alpha} (L) \right|^2.
 \label{alpha2beta}
\end{eqnarray}

Note that $P(\bar\nu_{\alpha} \rightarrow \bar\nu_{\beta})$ is related
to $P(\nu_{\alpha} \rightarrow \nu_{\beta})$ through $a \rightarrow
-a$ and $U \rightarrow U^{\ast} ({\rm i.e.\,} \delta \rightarrow
-\delta)$.
Similarly, we obtain $P(\nu_{\beta} \rightarrow \nu_{\alpha})$ from 
eq.(\ref{alpha2beta}) by replacing $\delta \rightarrow -\delta$,
$P(\bar\nu_{\beta} \rightarrow \bar\nu_{\alpha})$ by $a \rightarrow -a$.


Attributing both solar neutrino deficit and atmospheric neutrino 
anomaly to neutrino oscillation, we can assume
$a, \delta m^2_{21} \ll \delta m^2_{31}$.  The oscillation 
probabilities in this case can be considered by perturbation\cite{AKS}.
With the additional conditions
\begin{equation}
    \frac{aL}{2E}
    =
    1.93 \times 10^{-4} \cdot
    \left( \frac{\rho}{\rm g\,cm^{-3}} \right)
    \left( \frac{L}{\rm km} \right)
    \ll 1
    \label{AKScond1}
\end{equation}
and
\begin{equation}
    \frac{\delta m^{2}_{21} L}{2E}
    =
    2.53
    \frac{(\delta m^{2}_{21} / {\rm eV^{2}})(L / {\rm km})}{E / {\rm 
    GeV}} \ll 1,
    \label{AKScond2}
\end{equation}
the oscillation probabilities are calculated, e.g., as
\begin{eqnarray}
& &
 P(\nu_{\mu} \rightarrow \nu_{\rm e}; E, L)
=
 4 \sin^2 \frac{\delta m^2_{31} L}{4 E}
 c_{\phi}^2 s_{\phi}^2 s_{\psi}^2
 \left\{
  1 + \frac{a}{\delta m^2_{31}} \cdot 2 (1 - 2 s_{\phi}^2)
 \right\}
 \nonumber \\
&+&
 2 \frac{\delta m^2_{31} L}{2 E} \sin \frac{\delta m^2_{31} L}{2 E}
 c_{\phi}^2 s_{\phi} s_{\psi}
 \left\{
  - \frac{a}{\delta m^2_{31}} s_{\phi} s_{\psi} (1 - 2 s_{\phi}^2)
  +
 \frac{\delta m^2_{21}}{\delta m^2_{31}} s_{\omega}
    (-s_{\phi} s_{\psi} s_{\omega} + c_{\delta} c_{\psi} c_{\omega})
 \right\}
 \nonumber \\
&-&
 4 \frac{\delta m^2_{21} L}{2 E} \sin^2 \frac{\delta m^2_{31} L}{4 E}
 s_{\delta} c_{\phi}^2 s_{\phi} c_{\psi} s_{\psi} c_{\omega}
 s_{\omega}.
 \label{eq:AKSmu2e}
\end{eqnarray}
As stated, oscillation probabilities such as $P(\bar\nu_{\mu} 
\rightarrow \bar\nu_{\rm e})$, $P(\nu_{\rm e} \rightarrow \nu_{\mu})$
and $P(\bar\nu_{\rm e} \rightarrow \bar\nu_{\mu})$
are given from the above formula by some appropriate changes of the 
sign of $a$ and/or $\delta$.

The first condition (\ref{AKScond1}) of the approximation leads 
to a constraint for the baseline 
length of long-baseline experiments as
\begin{equation}
    L \ll
    1.72 \times 10^{3} {\rm km}
    \left( \frac{\rho}{3 {\rm g\,cm^{-3}}} \right)
    \label{eq:AKScond1'}
\end{equation}
The second condition (\ref{AKScond2}) gives the energy region where we 
can use the approximation,
\begin{equation}
    E
    \gg
    76.0 {\rm MeV}
    \left( \frac{\delta m^{2}_{21}}{10^{-4} {\rm eV^{2}}} \right)
    \left( \frac{L}{300 {\rm km}} \right).
    \label{eq:AKScond2'}
\end{equation}


\begin{figure}
 \unitlength=1cm
 \begin{picture}(15,18)
  \unitlength=1mm
  \centerline{
   \epsfysize=18cm
   \epsfbox{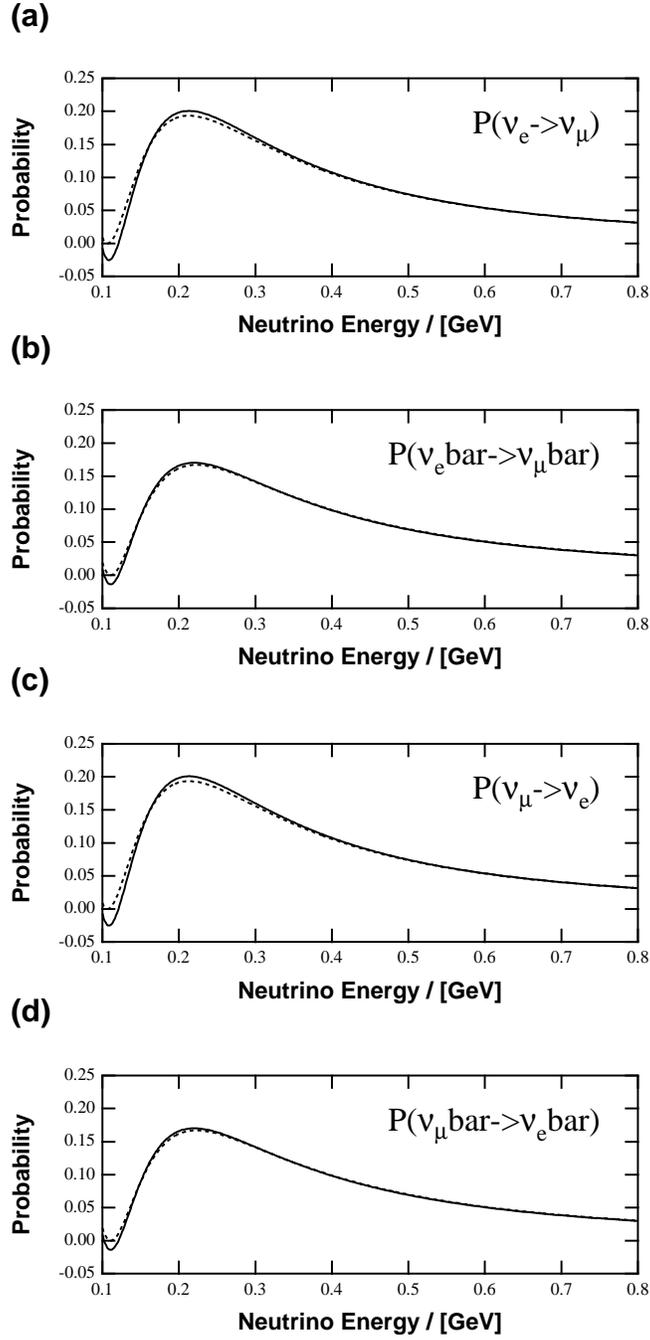}
  }
 \end{picture}
\caption{%
The approximated oscillation probabilities (solid lines)
compared with unapproximated ones (dashed lines).  Here the parameters
are taken as follows: $\delta m^{2}_{31} = 1.0 \times 10^{-3} {\rm
eV}^{2}, \delta m^{2}_{21} = 1.0 \times 10^{-4} {\rm eV}^{2}, \sin^{2}
\psi = 1/2, \sin^{2} \omega = 1/2, \sin^{2} \phi = 0.1, \sin \delta =
1; \rho = 2.5 {\rm g/cm^{3}}$ and $L = 300 {\rm km}$.
}
\label{fig1}
\end{figure}


We compare in Fig.\ref{fig1} the approximated oscillation
probabilities ((\ref{eq:AKSmu2e}) etc.)  with unapproximated ones to
show the validity of this approximation.  Here we set the baseline
length to be 300 km which corresponds to the distance between Tokai
and Kamioka.  Other parameters are taken from the region allowed by
present experiments\cite{Fogli}\footnote{Although the Chooz reactor
experiment have almost excluded $\sin^2\phi=0.1$\cite{chooz}, there
remains still small chance to take this value.} .  We see that the
approximation coincides with full calculation pretty well, and we are
safely able to use approximated formulas in the following.

\section{CP violation search in long baseline experiments}

\subsection{Magnitude of CP violation and matter effect}

The available neutrinos as an initial beam are $\nu_{\mu}$ and
$\bar\nu_{\mu}$ in the current long baseline
experiments\cite{K2K,Ferm}.  The ``CP violation'' gives the nonzero
difference of the oscillation probabilities between, e.g.,
$P(\nu_{\mu} \rightarrow \nu_{\rm e})$ and $ P(\bar\nu_{\mu}
\rightarrow \bar\nu_{\rm e})$\cite{AKS}.  This gives
\begin{eqnarray}
 P(\nu_{\mu} \rightarrow \nu_{\rm e}; L)
-
 P(\bar\nu_{\mu} \rightarrow \bar\nu_{\rm e}; L)
&=&
 16 \frac{a}{\delta m^2_{31}} \sin^2 \frac{\delta m^2_{31} L}{4 E}
 c_{\phi}^2 s_{\phi}^2 s_{\psi}^2 (1 - 2 s_{\phi}^2)
 \nonumber \\
&-&
 4 \frac{a L}{2 E} \sin \frac{\delta m^2_{31} L}{2 E}
 c_{\phi}^2 s_{\phi}^2 s_{\psi}^2 (1 - 2 s_{\phi}^2)
 \nonumber \\
&-&
 8 \frac{\delta m^2_{21} L}{2 E}
 \sin^2 \frac{\delta m^2_{31} L}{4 E}
 s_{\delta} c_{\phi}^2 s_{\phi} c_{\psi} s_{\psi} c_{\omega}
 s_{\omega}.
 \label{eq:CP}
\end{eqnarray}
The difference of these two, however, also includes matter effect, or
the fake CP violation, proportional to $a$.  We must somehow
distinguish these two to conclude the existence of CP violation as
discussed in ref.\cite{AKS}.

On the other hand, a muon ring enables to extract $\nu_{\rm e}$ and
$\bar\nu_{\rm e}$ beam.  It enables direct measurement of
pure CP violation through ``T violation'',
e.g., $P(\nu_{\mu} \rightarrow
\nu_{\rm e}) - P(\nu_{\rm e} \rightarrow \nu_{\mu})$ as
\begin{equation}
 P(\nu_{\mu} \rightarrow \nu_{\rm e}) -
 P(\nu_{\rm e} \rightarrow \nu_{\mu})
=
 - 8 \frac{\delta m^2_{21} L}{2 E}
 \sin^2 \frac{\delta m^2_{31} L}{4 E}
 s_{\delta} c_{\phi}^2 s_{\phi} c_{\psi} s_{\psi} c_{\omega}
 s_{\omega}.
 \label{eq:T}
\end{equation}
Note that this difference gives pure CP violation.

By measuring ``CPT violation'', e.g.
the difference between $P(\nu_{\mu} \rightarrow 
\nu_{\rm e})$ and $P(\bar\nu_{\rm e} \rightarrow \bar\nu_{\mu})$,
we can check the matter effect.
\begin{eqnarray}
 P(\nu_{\mu} \rightarrow \nu_{\rm e}; L)
-
 P(\bar\nu_{\rm e} \rightarrow \bar\nu_{\mu}; L)
&=&
 16 \frac{a}{\delta m^2_{31}} \sin^2 \frac{\delta m^2_{31} L}{4 E}
 c_{\phi}^2 s_{\phi}^2 s_{\psi}^2 (1 - 2 s_{\phi}^2)
 \nonumber \\
&-&
 4 \frac{a L}{2 E} \sin \frac{\delta m^2_{31} L}{2 E}
 c_{\phi}^2 s_{\phi}^2 s_{\psi}^2 (1 - 2 s_{\phi}^2)
 \nonumber \\
 \label{eq:CPT}
\end{eqnarray}
%


\begin{figure}
 \unitlength=1cm
 \begin{picture}(15,18)
  \unitlength=1mm
  \centerline{
   \epsfysize=18cm
   \epsfbox{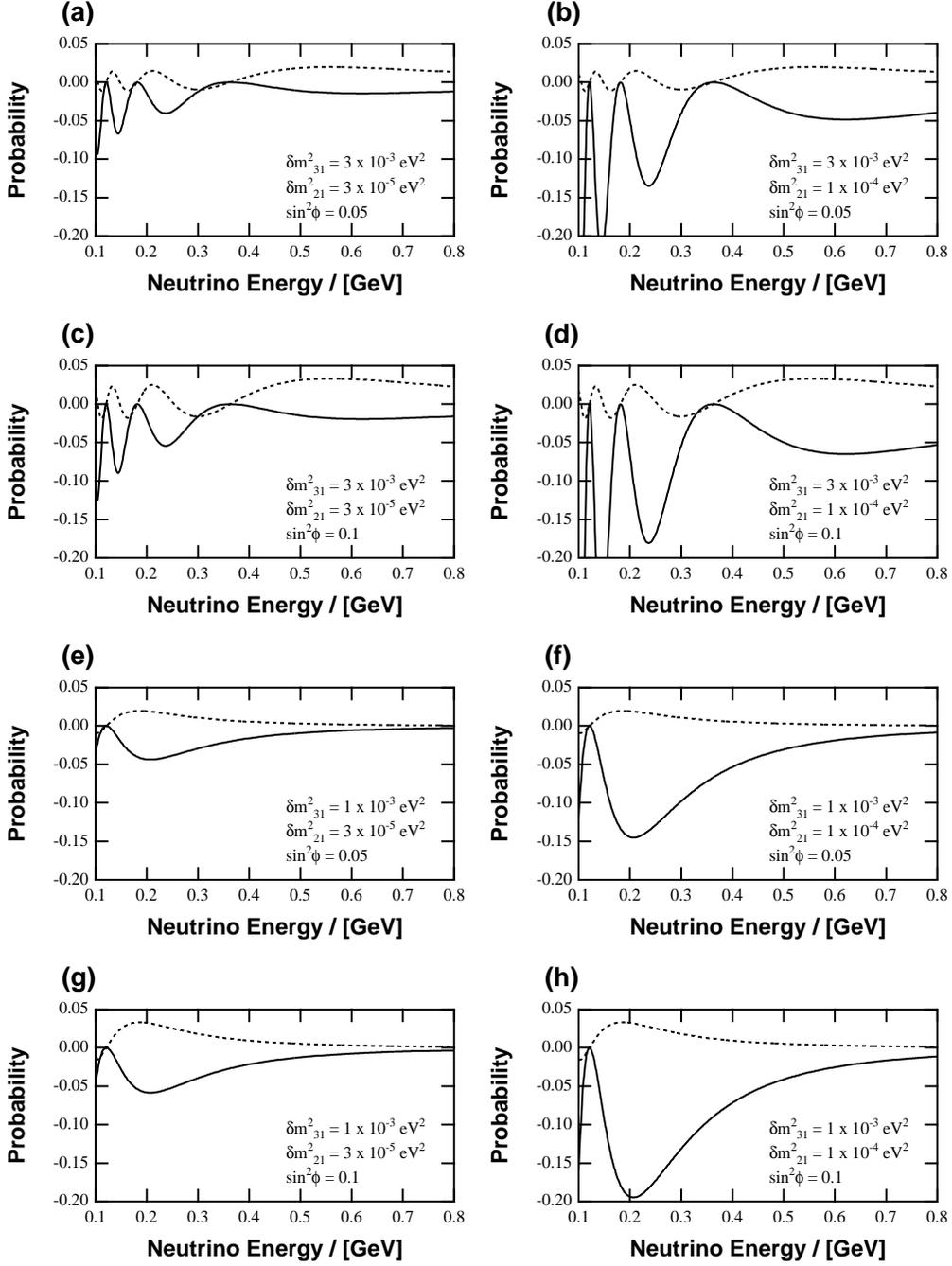}
  }
 \end{picture}
\caption{ Graphs of $P(\nu_{\mu} \rightarrow \nu_{\rm e}) - P(\nu_{\rm
e} \rightarrow \nu_{\mu})$ (solid lines; pure CP-violation effects) and
$P(\nu_{\mu} \rightarrow \nu_{\rm e}) - P(\bar\nu_{\rm e} \rightarrow
\bar\nu_{\mu})$ (dashed lines; matter effects) as functions of neutrino
energy.  Parameters not shown in the graphs are taken same as in
Fig.\ref{fig1}; $\sin^{2} \omega = 1/2, \sin^{2} \psi = 1/2, \sin \delta
= 1; \rho = {\rm g/cm^{3}}$ and $L = 300 {\rm km}$.} \label{fig2}
\end{figure}


We present in Fig.\ref{fig2} ``T-violation'' part (\ref{eq:T}) and
``CPT-violation'' part (\ref{eq:CPT}) for some parameters allowed by
the present experiments\cite{Fogli} with $\sin^{2}
\omega = 1/2$, $\sin^{2} \psi = 1/2$, $\sin \delta = 1$ fixed.  The
matter density is also fixed to the constant value $\rho = 2.5 {\rm
g/cm^{3}}$\cite{KS}.  Other parameters are taken as $\delta m^{2}_{31}
= 3 \times 10^{-3} {\rm eV}^{2}$ and $1 \times 10^{-3} {\rm eV}^{2}$,
$\delta m^{2}_{21} = 1 \times 10^{-4} {\rm eV}^{2}$ and $3 \times
10^{-5} {\rm eV}^{2}$.

``T-violation'' effect is proportional to $\delta m^{2}_{21} / \delta
m^{2}_{31}$ and, for $\phi \ll 1$, also to $\sin \phi$ as seen in
eq.(\ref{eq:T}) and Fig.\ref{fig2}.  Recalling that the energy of neutrino 
beam
is of several hundreds MeV, we see in Fig.\ref{fig2} that the
``T-violation'' effect amounts to at least about 5\%, hopefully
10$\sim$20\%.  This result gives hope to detect the pure leptonic CP
violation directly with the neutrino oscillation experiments.

The ``T violation'' is, however, less than 10\% in the case that
$\delta m^{2}_{21}$ is as small as $3 \times 10^{-5} {\rm eV}^{2}$
(see the left four graphs of Fig.\ref{fig2}).  In this case matter
effect is as large in magnitude as ``T violation'' and has an opposite
sign for $\sin\delta>0$ as seen in Fig.\ref{fig2}.  In such a case the
sum of the two, eq.(\ref{eq:CP}), is destructive and has even more
smaller magnitude than ``T violation'', thus the experiments will be
more difficult.  Thanks to $\nu_{\rm e}$ and $\bar\nu_{\rm e}$
available from low energy muon source, one can measure ``T
violation''.  This makes the measurement much easier.


\begin{figure}
 \unitlength=1cm
 \begin{picture}(15,18)
  \unitlength=1mm
  \centerline{
   \epsfysize=18cm
   \epsfbox{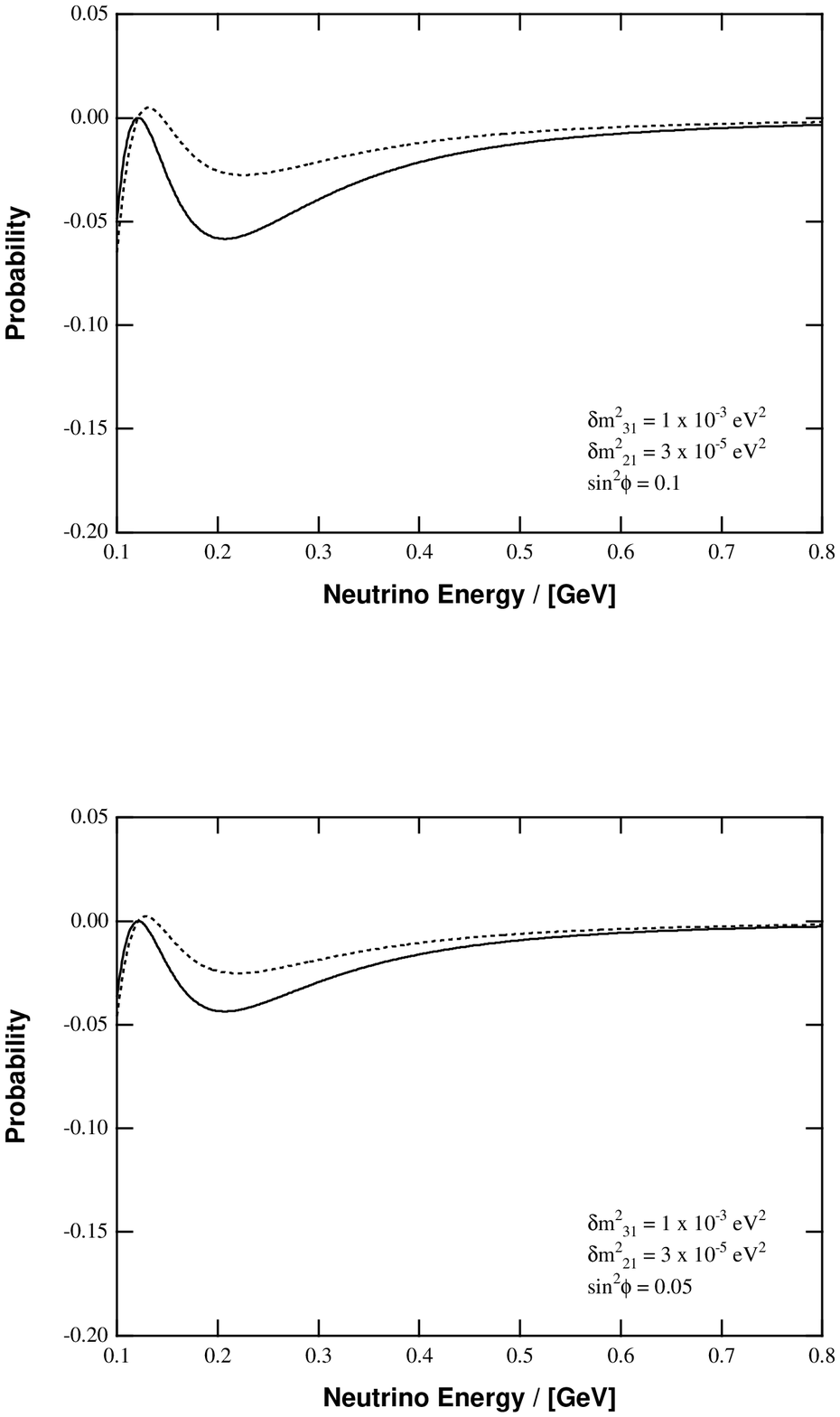}
  }
 \end{picture}
\caption{%
We compare the magnitudes of ``T-violation'' (eq.(\ref{eq:T}))
and the ``CP violation'' (eq.(\ref{eq:CP})) for some parameters. 
Parameters not shown in the graphs are taken same as Fig.\ref{fig2}.
}
\label{fig3}
\end{figure}


In Fig.\ref{fig3} we compare the magnitudes of ``T violation''
(eq.(\ref{eq:T})) and the ``CP violation'' (eq.(\ref{eq:CP})) for some
cases.  The peak value of ``T violation'' is almost twice larger than
that of ``CP violation''.  We consider that this is a major advantage of
the availability of the initial $\nu_{\rm e}(\bar\nu_{\rm e})$ beam.

\subsection{Estimation of statistical error in CP-violation searches}

Here we state that the energy range considered here is probably best
in view of statistical errors in order to observe CP violation effect. 
To this end let us estimate how $\delta P / \Delta P$ scales with $E$
and $L$, where $\delta P$ be statistical error of transition
probabilities such as $P(\nu_{\rm e} \rightarrow \nu_{\mu})$ and
$\Delta P = P(\nu_{\rm e} \rightarrow \nu_{\mu}) - P(\nu_{\mu}
\rightarrow \nu_{\rm e})$.  We denote in this section the transition
probabilities $P(\nu_{\alpha} \rightarrow \nu_{\beta}) (\alpha \ne
\beta)$ simply by $P$.  Suppose that $n$ neutrinos out of $N$ detected
neutrinos has changed its flavor.  With a number of decaying muons
fixed, the number of detected neutrinos $N$ are roughly proportional
to $E^3$, and hence $N \sim E^{3} L^{-2}$.  We estimate $\delta P$ as

\begin{eqnarray}
    \delta P &=&
    \delta \left( \frac{n}{N} \right)
    \nonumber \\
    &=&
    \frac{|N \delta n| + |n \delta N|}{N^{2}}
    \nonumber \\
    &=&
    \frac{|N \sqrt{NP}| + |NP \sqrt{N}|}{N^{2}}
    \nonumber \\
    &=&
    \frac{\sqrt{P} + P}{\sqrt{N}},
    \label{eq:deltaP}
\end{eqnarray}
where we used $\delta n = \sqrt{n}$, $\delta N = \sqrt{N}$ and $n = N
P$.  From eqs.(\ref{eq:AKSmu2e}), (\ref{eq:T}) and (\ref{eq:deltaP}),
we can estimate how $\delta P / \Delta P$ scales for $E$ with $L$
fixed.  We summarize the results in Table \ref{table:Escale}.  There
we see that $\delta P / \Delta P$ reaches minimum at the region $E
\sim \delta m^{2}_{31} L$. Note that 
this situation is quite different from that
for the transition probability $P$ itself.
\begin{table}
    \begin{center}
\begin{tabular}{|c||c|c|c|c|c|}
    \hline
    $E$ &  & $\delta m^{2}_{21} L$ & & $\delta m^{2}_{31} L $ & \\ 
    \hline \hline
    $P$ & ``const.'' &  & $1/E$ or ``const.'' & & $1/E^{2}$ \\
    \hline
    $\delta P$ & $1/E^{1.5}$ & & $1/E^{1.5} \sim 1/E^{2}$ & & $1/E^{2.5}$ \\
    \hline
    $\Delta P$ & ``const'' & & $1/E$ & & $1/E^{3}$ \\
    \hline \hline
    $\delta P / \Delta P$ & $1/E^{1.5}$ &  & $1/E^{0.5} \sim 1/E$ & & $E^{0.
5}$ 
    \\
    \hline
     & $\searrow$ & & $\searrow$ & minimum & $\nearrow$ \\
    \hline
\end{tabular}
\caption{The $E$-dependence of oscillation envelopes of some quantities
with $L$ fixed.  Here ``const.''  means that the oscillation envelope
of the quantity is independent of $E$.  $\delta P / \Delta P$ reaches
minimum at the region $E \sim \delta m^{2}_{31} L$.}
\label{table:Escale}
\end{center}
\end{table}

By a similar consideration one can obtain how $\delta P / \Delta P$ 
scales for $L$ with $E$ fixed.  The result for this case is shown in 
Table \ref{table:Lscale}.  We can see there that we should keep not 
too large $L$ so that the error $\delta P / \Delta P$ should not get 
large.
\begin{table}
    \begin{center}
\begin{tabular}{|c||c|c|c|c|c|}
    \hline
    $L$ &  &  $E/\delta m^{2}_{31} $ & & $E/\delta m^{2}_{21} $ & \\ 
    \hline \hline
    $P$ & $L^{2}$ &  & $L$ or ``const.'' & & ``const.'' \\
    \hline
    $\delta P$ & $L^{3}$ & & $L^{1.5} \sim L$ & & $L$ \\
    \hline
    $\Delta P$ & $L^{3}$ & & $L$ & & ``const.'' \\
    \hline \hline
    $\delta P / \Delta P$ & ``const.'' &  & $L^{0.5} \sim$ ``const.'' & & 
    $L$
    \\
    \hline
     & $\rightarrow$ &  & $\nearrow$ & & $\nearrow$ \\
    \hline
\end{tabular}
\caption{The $L$-dependence of oscillation envelopes of some quantities
with $E$ fixed.}
\label{table:Lscale}
\end{center}
\end{table}

We need a few hundreds MeV of neutrino energy to reach the threshold
energy of muon production reaction ${\rm N} + \nu_{\mu} \rightarrow
{\rm N} + \mu$, where N is nucleon.  We have also seen in
Table \ref{table:Escale} that the error comes to minimum at the region
$E \sim \delta m^{2}_{31} L$.  Considering these results, we conclude
that $E \sim$ (a few) $\times$ 100 MeV and $L \sim $ (a few) $\times$
100 km, which we have just considered in this paper, is the best
configuration to search CP violation in view of statistical error.

\section{Summary and conclusion}
We considered how large CP/T violation effects can be observed making
use of low-energy neutrino beam, inspired by PRISM. More than 10\%,
hopefully 20\% of the pure CP-violation effects may be observed within
the allowed region of present experiments.

We have also seen that in some case the pure CP-violation effects are
as small as the matter effect but have opposite sign.  In such a case
the ``CP violation'' gets smaller through the destructive sum of the
pure CP-violation effect and matter effect.  We pointed out that we
can avoid this difficulty by observing ``T-violation'' effect using
initial $\nu_{\rm e}$ beam.

We finally discussed that the configuration we have considered here,
$E \sim$ (a few) $\times$ 100 MeV and $L \sim$ (a few) $\times$ 100 km
is best to search lepton CP violation in terms of statistical error. 
It is thus worth making an effort to develop leptonic CP violation
search using neutrinos from low energy muons.


\section*{Acknowledgment}
The authors thank J. Arafune, Y. Kuno, Y. Mori and N. Sasao for useful
discussions.  One of the authors(J.S) is a research fellow of JPSJ.




\begin{thebibliography}{99}
    
\bibitem{Ga1} GALLEX Collaboration, W.~Hampel {\it et al.}, Phys.
    Lett. B {\bf 447}, 127 (1999) .
    
\bibitem{Ga2} SAGE Collaboration, J.~N.~Abdurashitov {\it et al.},
    astro-ph/9907113.
    
\bibitem{Kam} Kamiokande Collaboration, Y.~Suzuki, Nucl. Phys. B (Proc.
    Suppl.) {\bf 38}, 54 (1995).
    
\bibitem{Cl} Homestake Collaboration, B.~T.~Cleveland {\it et al.},
  Astrophys.  J. {\bf 496}, 505 (1998).


\bibitem{SolSK} Super-Kamiokande Collaboration, Y. Fukuda {\it et
  al.}, Phys. Rev. Lett. {\bf 82},1810 (1999), {\it ibid.} {\bf 82},
  2430 (1999).

\bibitem{AtmKam} Kamiokande Collaboration, K.~S.~Hirata {\it et al.},
  Phys. Lett.  {\bf B205},416 (1988); {\it ibid.}  {\bf B280},146 (1992);
  Y.~Fukuda {\it et al.}, Phys. Lett.  {\bf B335}, 237 (1994).

\bibitem{IMB} IMB Collaboration, D.~Casper {\it et al.},
  Phys. Rev. Lett. {\bf 66}, 2561 (1991);\\ R.~Becker-Szendy {\it et
  al.}, Phys. Rev.  {\bf D46}, 3720  (1992).

\bibitem{SOUDAN2} SOUDAN2 Collaboration, T.~Kafka, Nucl. Phys. B
  (Proc.  Suppl.) {\bf 35}, 427 (1994); M.~C.~Goodman, {\it ibid.} {\bf
  38}, 337 (1995); W.~W.~M.~Allison {\it et al.}, Phys. Lett. {\bf
  B391}, 491 (1997).

\bibitem{MACRO} MACRO Collaboration, M. Ambrosio {\it et al.},
Phys. Lett. {\bf B434}, 451 (1998). 

\bibitem{AtmSK} Super-Kamiokande Collaboration, Y. Fukuda {\it et
  al.}, Phys. Rev. Lett. {\bf 81}, 1562 (1998), Phys. Lett.  {\bf B433},
  9 (1998), Phys. Lett.  {\bf B436}, 33 (1998), Phys. Rev. Lett. {\bf
  82},2644 (1999).

\bibitem{LSND} LSND Collaboration, C.Athanassopoulos {\it et al.},
 Phys. Rev. Lett {\bf 77}, 3082 (1996); {\it ibid} {\bf 81},
1774 (1998).

\bibitem{FukugitaYanagida} For a review, M. Fukugita and T. Yanagida,
    in {\it Physics and Astrophysics of Neutrinos}, edited by M.
    Fukugita and A. Suzuki (Springer-Verlag, Tokyo, 1994).

\bibitem{KP} P. I. Krastev and S. T. Petcov, Phys. Lett. {\bf B205}, 
84 (1988).

\bibitem{Tanimoto} M. Tanimoto, Phys. Rev. {\bf D55}, 322 (1997);
    Prog. Theor. Phys.{\bf 97}, 901 (1997).
    
\bibitem{ArafuneJoe} J. Arafune and J. Sato, Phys. Rev.  {\bf D55},
    1653 (1997).

\bibitem{AKS} J. Arafune, M. Koike and J. Sato, 
    Phys. Rev. {\bf D56}, 3093 (1997).

\bibitem{MN} H. Minakata and H. Nunokawa, Phys. Rev. {\bf D57} 4403 (1998);
 Phys. Lett. {\bf B413}, 369 (1997).

\bibitem{BGG} M. Bilenky, C. Giunti, W. Grimus, Phys. Rev. {\bf D58},
 033001 (1998).

\bibitem{BGW} V. Barger, S. Geer and K. Whisnant, hep-ph/9906487.

\bibitem{Tanimoto2} M. Tanimoto, hep-ph/9906516; A. Kalliomki,
    J. Maalampi, M. Tanimoto, hep-ph/9909301.

\bibitem{Romanino} A. Romanino, hep-ph/9909425.

\bibitem{GH} A. De Rujula, M. B. Gavela and P. Hemandez,
    Nucl. Phys. {\bf B547}, 21 (1999); A. Donini, M. B. Gavela, P. 
    Hemandez and S. Rigolin, hep-ph/9909254.

\bibitem{Geer} S. Geer, Phys. Rev. {\bf D57}, 6989 (1998), erratum
    {\it ibid.} {\bf D59}, 039903 (1999).

\bibitem{Cabibbo} N. Cabibbo, Phys. Lett. {\bf B72}, 333 (1978).

\bibitem{BWP} V. Barger, K. Whisnant and R. J. N. Phillips, Phys. Rev. 
Lett. {\bf 45}, 2084 (1980).

\bibitem{PRISM} Y. Kuno and Y. Mori, Talk at the ICFA/ECFA Workshop
    ``Neutrino Factories based on Muon Storage Ring'', July 1999;
    Y. Kuno, in {\it Proceedings of the Workshop on High Intensity
    Secondary Beam with Phase Rotation}.

\bibitem{BilenkyPetcov} S. M. Bilenky and S. T. Petcov,
    Rev. Mod. Phys. {\bf 59}, 671 (1987).
    
\bibitem{Pakvasa} S. Pakvasa, in {\it High Energy Physics -- 1980},
    Proceedings of the 20th International Conference on High Energy
    Physics, Madison, Wisconsin, edited by L. Durand and L. Pondrom,
    AIP Conf. Proc. No. 68 (AIP, New York, 1981), Vol. 2, pp. 1164.

\bibitem{MNS} Z. Maki, M. Nakagawa and S. Sakata, Prog. Theor. Phys. 
{\bf 28}, 870 (1962).

\bibitem{ChauKeung} L. -L. Chau and W. -Y. Keung, Phys. Rev. Lett.
    {\bf 59}, 671 (1987).

\bibitem{KuoPantaleone} T. K. Kuo and J. Pantaleone, Phys. Lett. B
    {\bf 198}, 406 (1987).

\bibitem{Toshev} S. Toshev, Phys. Lett. B {\bf 226}, 335 (1989).
    
\bibitem{Wolf} L. Wolfenstein, Phys. Rev. {\bf D17}, 2369 (1978).

\bibitem{MS} S. P. Mikheev and A. Yu. Smirnov,
    Sov. J. Nucl. Phys. {\bf 42}, 913 (1985).

\bibitem{Earth} V. Barger, S. Pakvasa, R. J. N. Phillips and K.
    Whisnant, Phys.  Rev.  {\bf D22}, 2718 (1980).

\bibitem{K2K} K. Nishikawa, INS-Rep-924 (1992).
 
\bibitem{Ferm} S. Parke, Fermilab-Conf-93/056-T (1993),
    hep-ph/9304271.

\bibitem{Fogli} G. L. Fogli, E. Lisi, A. Marrone and G. Scioscia,
    Phys. Rev. D {\bf 59}, 033001 (1999).

\bibitem{chooz} M. Apollonio {\it et al.}, Phys. Lett. {\bf B420},
 397 (1998); hep-ex/9907037.

\bibitem{KS} M. Koike and J.Sato, Mod. Phys. Lett. {\bf A14},1297 (1999).

\end{thebibliography}
\end{document}